\documentclass{llncs}
\usepackage{graphicx}
\usepackage{color}
\usepackage{url}

\begin{document}

\title{
  Underspecified Scientific Claims in Nanopublications
}

\author{
  Tobias Kuhn and Michael Krauthammer
}

\institute{
  Department of Pathology,
  Yale University School of Medicine\smallskip\\
  \texttt{kuhntobias@gmail.com}, \texttt{michael.krauthammer@yale.edu}
}

\maketitle

\begin{abstract}
The application range of nanopublications --- small entities of scientific results in RDF representation --- could be greatly extended if complete formal representations are not mandatory. To that aim, we present an approach to represent and interlink scientific claims in an underspecified way, based on independent English sentences.
\end{abstract}

\section{Introduction}

This position paper introduces an approach to represent and interlink scientific statements with Semantic Web techniques, where these statements themselves do not necessarily have complete formal representations. To this aim, an extension of the concept of nanopublications is sketched. Nanopublications have been developed to make it easier to find, connect and curate core scientific statements and to determine their attribution, quality and provenance \cite{groth2010isu}. Small RDF-based data snippets --- i.e. nanopublications --- rather than classical narrative articles should be at the center of general scholarly communication \cite{mons2011naturegen}. Nanopublications are based on RDF extended with named graphs \cite{carroll2005www}.

There seem to be two possible types of nanopublications: they can represent claims or data. Data is directly observed from experiments or studies, whereas claims are obtained from generalizing from such data. The approach presented here has a clear focus on claims and not so much on data statements. ``Malaria is transmitted by mosquitoes'' \cite{groth2010isu} is a simple example of such a claim.

\section{Approach}

\begin{figure}[t]
\begin{center}
\includegraphics[width=0.98\textwidth]{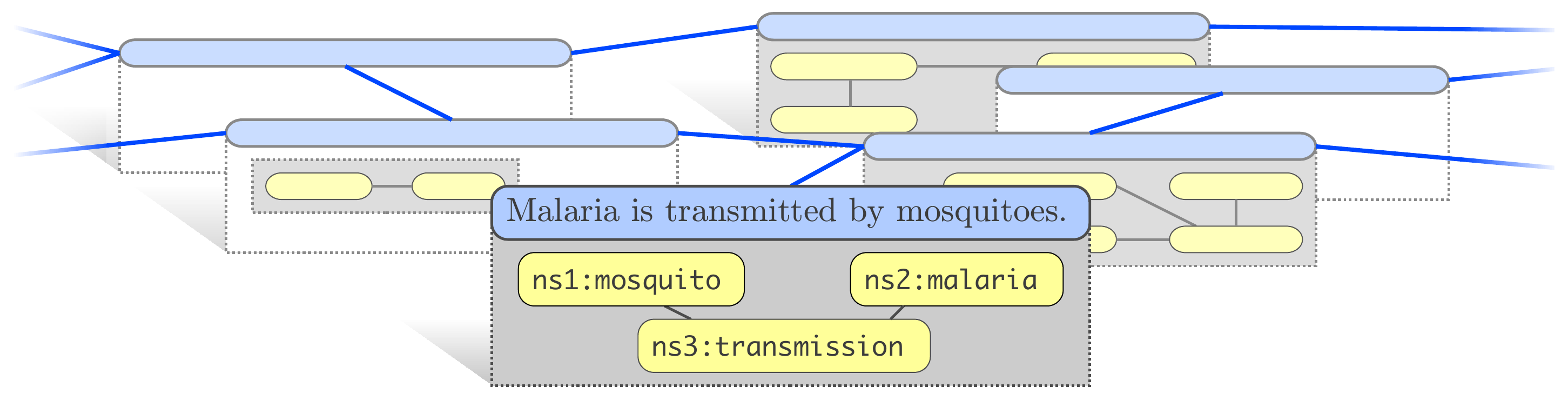}
\caption{Schematic representation of our model of scientific claims and their relations}
\label{fig:diagram}
\end{center}
\end{figure}

The proposed approach is based on the idea that any scientific claim can be broken down into small pieces of ``atomic'' claims, each of which can be represented as a relatively short independent sentence in English (or another natural language, possibly using highly technical vocabulary). Even though most claims found in scientific publications are probably more complex than ``malaria is transmitted by mosquitoes'', it seems reasonable to assume that they can be written down as independent sentences. By \emph{independent} we mean that the sentence can stand on its own and does not contain references like ``this behavior'' that refer to some surrounding text. Nanopublications follow the same basic idea, but require the claims to be fully formalized in RDF. We propose to extend nanopublications with English sentences, which are the central part of our model of scientific claims. Figure \ref{fig:diagram} shows a schematic representation of several claims according to our model. Each of the blue boxes contains an English sentence that represents the respective claim. Some claims have an additional formal representation in RDF (gray area), some do not (white area), and some are a mixture of the two (i.e. partial formalization). The important part is that all these claims, no matter whether formalized in RDF or not, can be interrelated and referenced, as indicated by the blue lines. These could be relations like ``{CLAIM1 contradicts CLAIM2}'' or ``PERSON agrees with CLAIM''. The white areas do not need to stay white forever: some of them might be filled with an RDF representation at a later point in time.

One could argue that any scientific claim can be represented in RDF in one way or another, given the appropriate vocabulary. In practice, however, the available vocabularies and ontologies are often not sufficient, especially for claims involving intended vagueness, modal concepts, temporal aspects, and novel ideas. RDF is extensible, but the development of accurate, useful and accepted models is a costly and slow process.
By dropping the restriction that all claims need full RDF representations, the application range of nanopublications can be greatly extended.

As a more realistic example, let us consider the following sentence from the abstract of a biomedical article (PMID 19109537):
\begin{quote}\small
[...] the risk of developing neurodegenerative disease in idiopathic REM sleep behavior disorder is substantial, with the majority of patients developing Parkinson disease and Lewy body dementia.
\end{quote}
These are the two core claims that can be extracted as independent sentences:
\begin{itemize}\small
\item The risk of developing neurodegenerative disease in idiopathic REM sleep behavior disorder is substantial.
\item The majority of patients with idiopathic REM sleep behavior disorder who develop a neurodegenerative disease develop Parkinson disease and Lewy body dementia.
\end{itemize}
To make these two sentences independent from each other, some parts have to be repeated. Still, the resulting sentences are reasonably short. The first one is a good example of vagueness in such claims (``substantial'').

\section{Integration}

Here, we sketch how the ideas described above could be integrated into the existing standards. As a first step, to be able to refer to statements like scientific claims even if they are not fully represented in RDF, we need URIs for such entire statements. We put forward the point of view that such a statement is simply a string of characters to be interpreted according to a certain language, like English or German. We use URIs instead of RDF string literals, because the latter cannot be used in subject position of RDF triples. Such a statement URI could be {\small\url{http://statements.org/en/Malaria+is+transmitted+by+mosquitoes}}.
Its semantics would be defined as all possible meanings that are given to it by the speakers of the respective language. This means that the authority behind such URIs (i.e. the fictitious \texttt{\small statements.org} in the given example) would not need to approve new statements, but everybody could make up such URIs and immediately use them. As a next step, we can integrate them in nanopublications.

The core part of a standard nanopublication is an \emph{assertion} in the form of a named graph:
\begin{quote}
\small
\vspace{-0.1ex}
\begin{verbatim}
<> {
  :Pub1 np:hasAssertion :Pub1_Assertion .
  ...
}
:Pub1_Assertion { ... }
\end{verbatim}
\vspace{-0.1ex}
\end{quote}
The curly brackets after \texttt{\small :Pub1\_Assertion} would contain the actual assertion in the form of a set of RDF triples. To allow for underspecified assertions, we have to use a slightly more complex structure. With our approach, assertions consist of two subgraphs: a head and a body, where the body represents the actual (possibly unknown) formal representation:
\begin{quote}
\small
\vspace{-0.1ex}
\begin{verbatim}
<> {
  :Pub1 np:hasAssertion :Pub1_Assertion .
  :Pub1_Assertion np:containsGraph :Pub1_Assertion_Head .
  :Pub1_Assertion np:containsGraph :Pub1_Assertion_Body .
  ...
}
\end{verbatim}
\vspace{-0.1ex}
\end{quote}
The head part is used to refer to different representations of the given assertion, such as the formal representation in the form of a named RDF graph or a natural representation in the form of an English sentence encoded in a URI:
\begin{quote}
\small
\vspace{-0.1ex}
\begin{verbatim}
:Pub1_Assertion_Head {
  :Pub1_Assertion
      st:asSentence st:en/Malaria+is+transmitted+by+mosquitoes ;
      st:asFormula :Pub1_Assertion_Body .
}
\end{verbatim}
\vspace{-0.1ex}
\end{quote}
We can --- but we are not obliged to --- add a formalization of the given claim with {\small\verb-:Pub1_Assertion_Body { ... }-}. Partial representations can be defined in a straightforward way with the help of subgraphs. Overall, this approach allows for defining nanopublications for virtually any possible scientific claim. Even claims that cannot be formalized in RDF can be included in the Semantic Web.

\section{Discussion}

There exist approaches like GeneRIF,\footnote{\url{http://www.ncbi.nlm.nih.gov/gene/about-generif}} which is based on a similar idea but is restricted to a very specific domain (gene functions). Our approach is much more general and could subsume such specific solutions.

The approach sketched above in a certain sense uses Semantic Web techniques on a higher level than usual. Instead of representing relations between entities of the real world, we relate \emph{statements} about the real world to other statements or entities. While such relations are no less fuzzy at this higher level than certain lower level relations, it is possible at the higher level to come up with a model that covers virtually all possible scientific claims. Many existing approaches based on RDF use this kind of higher level (e.g. provenance data for reified RDF triples), but they typically require the lower level to be spelled out too. We try to advocate the idea that we can describe things at the higher level without being specific about the lower one. Of course, it is always better to have RDF representations for both levels, but having just the higher one is better than nothing in cases where the lower level cannot be practically formalized (which might very well be the majority of cases).


Even though we only presented examples in English, our approach is inherently multilingual, as claims can be verbalized in different languages. Furthermore, instead of using unrestricted language, scientific claims could be expressed in a controlled natural language \cite{wyner2009cnlmain}, in which case RDF representations could be automatically generated (depending on the used controlled natural language). Previous work indicates that this could be feasible for at least certain types of scientific claims \cite{kuhn2006dils}.

We hope to be able to present a concrete proposal for underspecified nanopublications in the near future. We also plan to evaluate our approach by assessing scientific claims of existing publications.

\bibliography{nanopub}
\bibliographystyle{plain}

\end{document}